\documentclass{aa}

\usepackage{psfig}

\newcommand{\rsol}{{R$_{\odot}$}}

\newcommand{\ks}{km s$^{-1}$}

\newcommand{\M}{{\sc 2mass}}

\newcommand{\DE}{{\sc denis}}

\include{psfig.sty}

\begin{document}


\thesaurus{03(08.02.2, 08.04.1, 08.09.2, 11.13.1, 12.04.3)}
\title{The LMC eclipsing binary HV 2274 revisited}

\author{M.A.T. Groenewegen\inst{1,3} \and M. Salaris\inst{2,1} }

\offprints{Martin Groenewegen, e-mail: mgroenew@eso.org}

\institute {
Max-Planck-Institut f\"ur Astrophysik, 
Karl-Schwarzschild-Stra{\ss}e 1, D-85740 Garching, Germany
\and 
Astrophysics Research Institute, Liverpool John Moores
University, Twelve Quays House, Egerton Wharf, Birkenhead CH41 1LD, UK
\and
current address: European Southern Observatory, EIS-team, 
Karl-Schwarzschild-Stra{\ss}e 2, D-85740 Garching, Germany
}

\date{received,  accepted}


\maketitle

\begin{abstract}

We reanalyse the UV/optical spectrum and optical broad-band data of
the eclipsing binary HV 2274 in the LMC, and derive its distance
following the method given by Guinan et al.~(1998a,b) of fitting
theoretical spectra to the stars' UV/optical spectrum plus optical
photometry.  We describe the method in detail, pointing out the
various assumptions that have to be made; moreover, we discuss the
systematic effects of using different sets of model atmospheres and
different sets of optical photometric data. It turns out that
different selections of the photometric data, the set of model
atmospheres and the constraints on the value of the ratio of selective
to total extinction in the $V$-band, result in a 25\% range in
distances (although some of these models have a large ${\chi}^2$).

For our best choice of these quantities the derived value for the
reddening to HV 2274 is $E(B-V)$ = 0.103 $\pm$ 0.007, and the
de-reddened distance modulus is DM = 18.46 $\pm$ 0.06; the DM to
the center of the LMC is found to be 18.42 $\pm$ 0.07. This is
significantly larger than the DM of 18.30 $\pm$ 0.07 derived by Guinan
et al. (1998a).


\end{abstract}

\section{Introduction} 

The distance to the Large Magellanic Cloud (LMC) is a fundamental step
in the cosmological distance ladder, since the extragalactic distance
scale is usually determined with respect to the LMC distance. In fact,
both the {\sc hst} $H_0$ Key Project (Kennicutt et al. 1995, Freedman
et al. 1999) and the Supernovae Calibration Team (Saha et al. 1999)
fix the zero-point of the cosmological distance scale assuming a
de-reddened LMC distance modulus (DM) of 18.50; in the case of the {\sc
hst} $H_0$ Key Project the adopted uncertainty of $\pm$ 0.13 on the
LMC distance modulus represents the largest contribution to their
systematic error budget.

In recent years various methods have yielded LMC distance moduli
showing a remarkable spread, ranging from DM = 18.07 (Udalski et
al. 1998a) to DM = 18.70 (Feast \& Catchpole 1997) -- see for example
the review by Gibson (1999)  or Feast (2001). This uncertainty
alone on the LMC distance causes an indetermination by $\sim$20\% on
the value of the Hubble constant.

A method that, at present, seems to support the short distance scale,
involves the analysis of the light-curve, radial velocity curve and
UV/optical spectrum of the detached eclipsing binary HV 2274 in the
LMC.  It is considered (see, e.g., Gibson 1999) to be one of the most
promising techniques to derive a precise distance to the LMC, and it
is based on a very elegant idea. From the analysis of the radial
velocity and light-curve one obtains very accurate values for the
masses and radii of the two binary components, as well as for the
ratio of the effective temperatures. Fitting the UV/optical spectrum
with model atmospheres one obtains the reddening, effective
temperature and distance. This method, as already stressed, has been
put forward as {\sl the} way to obtain a very accurate DM to the LMC,
in particular when more systems will be analysed.

Guinan et al. (1998a; hereafter G98a) found DM = 18.30 $\pm$ 0.07 (the
derived distance modulus to HV 2274 was 18.35 $\pm$ 0.07. A geometric
correction has been then applied to obtain the distance to the center
of the LMC) after applying this technique.  More recently Nelson et
al.~(2000) corrected the G98a value after re-determining the reddening
toward HV 2274; they obtained DM = 18.40 $\pm$ 0.07, a value only
marginally in agreement with the long distance. However, they did not
apply the method by G98a, deriving the distance correction only in an
indirect way.

Because of the relevance of the method employed by G98a for deriving
the LMC distance and in light of the recent claims by Nelson et
al.~(2000) for a longer LMC distance, we want in the present paper to
re-analyse the fitting procedure of the UV/optical spectrum, and
carefully point out the uncertainties of this method. Unfortunately,
in their 4-page {\sl Letter}, G98a could not present all the intricate
details that {\sl are} involved in the application of this method, but
which should be pointed out to the scientific community in order to
judge the strengths and weaknesses of this technique (also see
Feast 2001).  We will also study the sensitivity of the derived
distance to both the set of model atmospheres and of optical
photometric data employed in the fitting procedure.

In Sect.~2 we present an historical overview of observations and
studies related to the HV 2274 distance. In Sect.~3 and 4, we discuss,
respectively, the observational data and the method  
employed for the distance determination. Results are presented 
in Sect.~5, while a discussion follows in Sect.~6.

\section{Historical overview}

Watson et al. (1992) presented Johnson $BV$ and Cousins
$I$ CCD photometry of HV 2274. The estimated out-of-eclipse magnitudes
are $V$ = 14.16, $(B-V) = -0.18$ and $(V-I) = -0.08$ with
uncertainties of ``not as large as 0.1 mag''. 



Guinan et al.~(1998b - hereafter G98b) discussed new observational
data of HV 2274. They took IUE UV (1200-3000 \AA) spectra and
fitted them with Kurucz ATLAS 9 model atmospheres.
They also obtained UV spectra (1150-4820 \AA) with the HST/FOS
spectrograph, and medium resolution spectra with the HST/GHRS to
obtain the radial velocity curve. The radial velocity curve and light
curve (from Watson et al. 1992) were analysed to provide accurate
determinations of (amongst others) the ratio of the effective
temperatures and the gravities. Then, using the method outlined below,
of fitting the FOS spectrum (plus $(B-V)$, from, in this case, Watson
et al.) with model atmospheres taking into account the effect of
extinction and using the constraints from the radial velocity and
light-curve solution, they derived the micro-turbulent velocity
(assumed equal for the 2 components), $E(B-V)$ and coefficients
describing the UV extinction in the HV 2274 line-of-sight, the
effective temperature of the primary, the metallicity (assumed equal
for the 2 components), and the distance (see Table~2). Considering the
fact that HV 2274 was estimated by them to be 1100 pc behind the
center of the LMC, they derived a DM = 18.44 $\pm$ 0.07.

Then, two papers appeared in the same issue of the {\sc Astrophysical
Journal}, Udalski et al. (1998b; hereafter UPW98) and G98a. The former
was a reaction to a {\sl pre-print} version of G98a
(astro-ph/9809132v1; which quoted $E(B-V) = 0.083 \pm 0.006$, and a DM
to the binary of 18.47 $\pm$ 0.07) and presented out-of-eclipse {\sc
ogle-ii} $UBVI$ photometry ($V = 14.16 \pm 0.02$, $(B-V) = -0.129 \pm
0.015$, $(V-I) = -0.125 \pm 0.015$, $(U-B) = -0.905 \pm 0.04$) of HV
2274.  UPW98 derived the reddening to HV 2274 from $(B-V)$ and
$(U-B)$, using the colours of unreddened Galactic B-stars, and
assuming the mean LMC reddening line of $E(U-B)/E(B-V) = 0.76$. They
derived $E(B-V) = 0.149 \pm 0.015$. They did not repeat the fitting
procedure in G98a and G98b, but noted that in the reddest part of the
spectrum, near 4800 \AA, the reddening is proportional to 3.8 $E(B-V)$
(Udalski et al. hence implicitly assumed a Galactic type reddening
curve), and derived a distance modulus that was shorter by (3.8
$\times (0.149 - 0.083) = 0.25$ mag, or DM = 18.22 $\pm$ 0.13. They
chose not to apply a geometrical correction.

The published version of G98a already considered a preprint version of
UPW98, and fitted the FOS spectrum plus $B,V$ data, not from Watson et
al. (1992 - as was done in G98b), but already with that from
UPW98. They derived $E(B-V) = 0.120 \pm 0.009$ and DM = 18.35 $\pm$
0.07 to HV 2274. This already indicates the sensitivity of the
solution to the adopted photometry that is included in the fitting
procedure.  We remark here that, as mentioned in G98a, it is necessary
to include in the fitting procedure photometric data in the wavelength
region between 4400 and 5500 \AA\ (hence the use of $(B-V)$ data),
otherwise a possible degeneracy between the parameters determining the
reddening law and $E(B-V)$ does exist.


Nelson et al.~(2000) noted the uncertainty in obtaining ``standard''
$U$ and $B$ photometry, and criticized UPW98 for using a non-standard
$U$ filter and few calibration observations. They obtained $UBV$
photometry of stars in a field around the binary. For HV 2274 they
obtained $V = 14.20 \pm 0.006$, $(B-V) = -0.172 \pm 0.013$, $(U-B) =
-0.793 \pm 0.027$. The errors came from the rms residuals in the
transformation from instrumental magnitudes to the standard system.
These colours differ significantly from those of UPW98. Nelson et
al.~(2000) ascribed this mainly to a difference in the $U$-filter,
and, likely, to a smaller extent, to differences in the
$B$-filter. They then used, similarly to UPW98, $(B-V)$ and $(U-B)$
colours to derive a reddening to HV 2274 of $E(B-V) = 0.088 \pm
0.025$. Without re-doing the analysis of G98a,b they noted that this
reddening is almost identical to that in G98b and therefore they
suggested that a DM of 18.40 $\pm$ 0.07 to the center of the LMC is
the appropriate one.

\section{The observational data}

The calibrated out-of-eclipse FOS spectrum consists of four
individual spectra, covering four adjacent wavelength regions
spanning the range between approximately 1200 and 4800 \AA. The
calibrated spectra have been taken from the {\sc hst} archive
and provide with observed wavelength, flux (in erg s$^{-1}$ cm$^{-2}$
\AA$^{-1}$), and the formal error on the flux. The observed
wavelengths are transformed to rest wavelengths using the measured
radial velocity of the binary system of +312 ($\pm$ 4) \ks\ (see Ribas
et al. 2000). It is not clear if this correction was applied by G98a,b
or not. We find that we obtain a significantly lower $\chi^2$
including this correction. On the other hand, the resolution of the
model atmospheres used is such that it is not necessary to consider
the fact that the 2 binary components have different radial
velocities, due to their orbital motion.

Before merging the data of different wavelength regions we have
compared the fluxes (with errors) in the overlap regions. We found no
reason to scale the different spectra, as possible scaling factors are
less than 1\%. This possible error is considered later-on in the
final error budget.

The final merged spectrum has been rebinned to the wavelength grid of
the model atmospheres we will use (see next Section), which are at a
lower resolution than the FOS spectrum. This step was also performed
by G98a,b. For each wavelength point ${\lambda}_{i}$ in the model
atmosphere we calculate the weighted mean and error of the flux of the
FOS spectrum between $0.5 \times ({\lambda}_{i-1} + {\lambda}_{i})$
and $0.5 \times ({\lambda}_{i} + {\lambda}_{i+1})$, using linear
interpolation to estimate the flux and error at the begin- and
end-point. Later on we will argue that these formal errors are likely
to be underestimates of the true errors.

As a last step one has to remove data points that are affected by
interstellar lines (as was done by G98a,b). In fact, the wavelengths
of these points can be read from Fig.~2 in G98a, but also stand out
clearly as deviations between the model fit (see below) and the
observations. After removing these points the final spectrum used in
the model fitting contains 254 data points between 1145 and 4790 \AA.

The other observational data needed to apply the method described in
G98a,b are broad-band photometry. G98b used $(B-V)$ from Watson et
al. (1992), while G98a used $(B-V)$ from UPW98, which
has a much smaller error.
There is also $(V-I)$ data available from UPW98,
which was not used by G98a as additional constraint, but will be used
by us together with the UPW98 $(B-V)$.  The fact that
G98a did not make use of the $(V-I)$ data is surprising as one of the
problems in the fitting procedure noted by them is a degeneracy
between the parameters determining the reddening law, and $E(B-V)$. An
additional constraint at longer wavelengths might therefore be very
helpful.

To summarise, the reference observational data we will use consist of
256 observational data points, 254 coming from the binned FOS spectrum
after removing points that are contaminated by interstellar lines,
$(B-V)$ and $(V-I)$.

Recently, the teams representing the \M\ $JHK_{\rm s}$ survey
(Beichman et al. 1998) and the \DE\ $IJK_{\rm s}$ survey (Epchtein et
al. 1999, Cioni et al. 2000) released data that contain the Magellanic
Cloud area.  HV 2274 is not in the \DE\ database, but it is in the \M\
survey. The position quoted is (J2000): RA= 5h 02m 40.74s, Dec =
$-68$d 24m 21.46s, with magnitudes $J = 15.152 \pm 0.057$, $H = 15.168
\pm 0.107$ and $K_{\rm s} = 15.520 \pm 0.238$. The \M\ User Support \&
Help Desk kindly checked their database to communicate the time of
observation to be 1998 Oct 25, 07h 27m 30.48s UT. This corresponds to
JD = 245 1111.689. These data can not be used as constraint as it will
be argued later that this observation happens to almost coincide with
a primary eclipse, and so can not be used to constrain the parameters
as determined from the out-of-eclipse FOS spectrum and $UBVI$
photometry.

\section{The method}

The method employed is essentially identical to that used by
G98a,b. Using the constraints that come from the (independent)
modelling of the radial velocity and light-curve, model atmospheres 
are fitted to the (binned) UV FOS spectrum and broad-band data. 

In more detail, the total out-of-eclipse flux of the binary at each
wavelength bin can be written as (see G98a):
\begin{displaymath}
f_{\rm model} (\lambda) = \left( \frac{r_{\rm A}}{d}\right)^2
[F_{\lambda}^{\rm A} + (r_{\rm B}/r_{\rm A})^2 F_{\lambda}^{\rm B}]
\end{displaymath}
\begin{equation}
\hspace{3cm}    \times 10^{-0.4 \; E(B-V)\;[k(\lambda-V) + R]}
\end{equation}
where $r_{\rm A}$ and $r_{\rm B}$ are the radii of, respectively, the
primary and secondary component, $d$ is the distance,
$F_{\lambda}^{\rm A}$ and $F_{\lambda}^{\rm B}$ the emergent fluxes
from the two components, $R$ is the ratio of selective to total
reddening in $V$, $E(B-V)$ is the colour excess and $k(\lambda-V)$ is
the normalised extinction curve defined as $\frac{ E(\lambda-V)
}{E(B-V)}$.

The model atmospheres we employed are described below; in general they
can be characterised by four parameters: effective temperature,
gravity, metallicity and micro-turbulent velocity. The micro-turbulent
velocity is fixed at 2 \ks, following the result of G98a (see below)
and is assumed to be the same for both components. The gravity of the
two components, the ratio of the stellar radii and the ratio of the
effective temperatures follow from the radial velocity and light-curve
analysis (see G98a, summarized in Table~1). Furthermore, it is assumed
that both components have the same metallicity. This leaves four
parameters: the effective temperature of the primary, the metallicity,
$E(B-V)$ and the scaling factor $\left(\frac{r_{\rm A}}{d}\right)^2$
from which the distance is derived ($d$ [kpc] = 70.26/$\sqrt{
\left(\left(\frac{r_{\rm A}}{d}\right) ^2 \; 10^{23} \right) }$ for
$r_{\rm A}$ = 9.86 \rsol, see Table~1).  The inclusion of the
extinction curve in Eq.~(1) is an essential ingredient in the
procedure, and is discussed in detail below. It involves in our case
five additional free parameters.

The procedure is as follows. By considering the constraints listed
in Tab.~1 the nine free parameters are varied so that the following
function is minimised:
\begin{displaymath}
\chi^2 = \left( \frac{(B-V)_{\rm model} - (B-V)_{\rm obs}}{\sigma_{\rm
B-V}}\right)^2 
\end{displaymath}
\begin{displaymath}
\hspace{1.5cm} + \left( \frac{(V-I)_{\rm model} - (V-I)_{\rm
obs}}{\sigma_{\rm V-I}}\right)^2
\end{displaymath}
\begin{equation}
\hspace{2.5cm} + \sum \; \left( \frac{\log f_{\rm model} (\lambda) -
\log f_{\rm obs} (\lambda)}{\sigma_{\rm obs} (\lambda)}\right)^2
\end{equation}
We use a Levenberg-Marquardt non-linear least-squares method (Press et
al. 1992). The method makes use of the derivatives of $\chi^2$ with
respect to the free parameters, which are calculated
numerically\footnote{The derivatives are calculated from
\begin{displaymath}
f^{\prime} = \frac{ f(x+h)-f(x-h)}{2\,h},
\end{displaymath}
where $h$ = $\epsilon \;\; x$. Following the considerations in Press
et al. (1992) and numerical experiments, we choose $\epsilon = 0.003$
for the variable effective temperature, $\epsilon = 0.03$ for the
variables metallicity and $c_4$, and $\epsilon = 0.01$ for all
others.}. The errors in the parameters are calculated from the
square-root of the diagonal elements of the covariance matrix (see
Press et al. 1992).

Before going on discussing the model atmospheres and extinction law
adopted in the fitting procedure, we want to comment briefly about the
parameters listed in Table 1. They are derived either directly from
the light curve analysis (e.g., $T_{\rm B}/T_{\rm A}$) or by combining
the results from the light curve analysis with the results from the
radial velocity curve solution. In particular, the value of $r_{\rm
A}$ is obtained from $r_{\rm A}=r_{\rm f}a$, where $a$ is the orbital
semimajor axis derived from the radial velocity curve, and $r_{\rm f}$
the fractional radius derived from the light-curve solution.  When
performing the light-curve analysis, apart from the need to assume
some value for the albedos and gravity-darkening exponents, it is
necessary to employ even at this stage theoretical model atmospheres
(see, e.g. the discussion in Milone et al.~1992), from which, for
example, limb-darkening coefficients are derived; in the case of HV
2274, G98ab used (according to Ribas et al.~2000) a version of the
Wilson-Devinney (1971) program that includes the Kurucz ATLAS 9 (see
next subsection) model atmosphere routine developed by Milone et
al.~(1992, 1994).  The errors quoted in Table 1 are formal errors
obtained from the given set of assumptions made by the authors and
from the observational errors.  It is dificult to assess the
uncertainty on the parameters estimated from the light curve analysis
due to uncertainties on model atmospheres. Milone et al.~(1992)
discussed in detail the case of the eclipsing binary AI Phoenicis;
even if the two components of this system are much colder objects than
HV 2274, it can be instructive to notice that by changing the
underlying model atmospheres (black body, Carbon-Gingerich~1969
models, Kurucz~1979 models, Kurucz~1979 models corrected for the
missing ultraviolet opacity) and band passes employed in the light
curve analysis, the value of the fractional radii of both components
are basically unchanged, while the ratio $T_{\rm B}/T_{\rm A}$ varies
at most by $\sim$1.6\%.

\begin{table}
\caption[]{Constraints from the binary astrometric analysis}

\begin{tabular}{lll}  \hline
Quantity  & Symbol & Value     \\ \hline
Effective temperature ratio & $T_{\rm B}/T_{\rm A}$   & 1.005 $\pm$ 0.005 \\
Gravity primary             & $\log g_{\rm A}$        & 3.536 $\pm$ 0.027 \\
Gravity secondary           & $\log g_{\rm B}$        & 3.585 $\pm$ 0.029 \\
Radius primary              & $ r_{\rm A}$            & 9.86 $\pm$ 0.24 \rsol\\
Radii ratio               & $(r_{\rm B}/r_{\rm A})^2$ & 0.842 $\pm$ 0.019 \\
\hline
\end{tabular}
\end{table}

\subsection{Model atmospheres}

We will consider three sets of model atmospheres: (1) the ATLAS 9
models by R.L. Kurucz taken from his homepage (Kurucz 2000); (2) model
atmospheres calculated by K. Butler (2000, private communication;
Butler et al., in preparation); (3) model atmospheres calculated by
I. Hubeny (2000, private communication) using the {\sc tlusty} code
(Hubeny 1988; Hubeny \& Lanz 1992; Hubeny et al. 1994, Hubeny \& Lanz
1995).

These codes are not identical. The ATLAS9 code assumes LTE and uses an
iron abundance of $A$(Fe) = 7.67 (on a scale where $\log H$ = 12).
There is an offset by 0.16 dex between this value 
and the recommended value of 7.51 ($\pm$ 0.01) by Grevesse \& Noels (1993).
The {\sc tlusty} code is a fully line-blanketed NLTE model, with
improved continuum opacities with respect to Kurucz (Hubeny, private
communication), and uses $A$(Fe) = 7.50.  Unfortunately, the
wavelength coverage is only up to 8200 \AA\ and therefore does not
cover the entire wavelength range of the $I$-band. In the code
used by Butler, hydrogen and helium are in NLTE, while CNO, Si, Mg and
the iron group metals are in LTE. The iron abundance used is $A$(Fe) =
7.46. There are also differences in the abundances of the other
metals, but these are generally smaller.

We considered model atmospheres for the following combination of
parameters: $T_{\rm eff}$ = 22~000 and 24~000 K, $\log g$ = 3.5 and
4.0, and metallicity [m/H] = $-1.0$, $-0.5$, $-0.3$ and $-0.0$.  We
had access also to ATLAS 9 models with different values of the
micro-turbulent velocity, and we left it as a free parameter in some
tests we performed, aimed at deriving the sensitivity of $E(B-V)$,
distance and $T_{\rm eff}$ on the value of the adopted micro-turbulent
velocity.  No significant differences were found with respect to the
case of fixing this parameter at 2 \ks, and from now on we will always
refer to models computed with this value of the micro-turbulent
velocity. G98a left this as a free parameter in their fit and derived
1.9 $\pm$ 0.7 \ks.

In the fitting procedure it is necessary to interpolate among 12
models (2 temperatures $\times$ 2 gravities $\times$ 3 metallicities,
picked among the 4 metallicities available). The interpolation is done
in $\log$ flux, linearly in $T_{\rm eff}$ and $\log g$, and
quadratically in metallicity. After the interpolated model is created,
a correction is made to enforce flux conservation, which typically
amounts to less than 0.5\%. The Butler and {\sc tlusty} models are
calculated at a higher resolution, and are rebinned to the exact
wavelength grid of the Kurucz models.

Although no details are explicitly given in G98a, they refer to
Fitzpatrick \& Massa (1999) were it is explained that they used the
ATLAS 9 code to calculate an extended grid of model
atmospheres, for different micro-turbulent velocities. They
interpolate quadratically in metallicity and micro-turbulent
velocity, and linearly in $\log g$ and $\log T_{\rm eff}$. Fitzpatrick
\& Massa (1999) state that they use the three metallicities closest to
the desired one, which should have been [m/H] = $-1.0$, $-0.5$ and
$0.0$ in their case.


\subsection{Theoretical colours}

From the theoretically predicted observed flux distribution we also
have to calculate broad-band colours and magnitudes to minimize the
expression given in Eq.~2. In general a magnitude in a generic
bandpass can be written as:
\begin{equation}
 m = ZP -2.5 \times \log \left(  \frac{\int f_{\lambda} R_{\lambda} 
d \lambda}{\int R_{\lambda} d \lambda} \right),
\end{equation}
where $ZP$ is the zero point, $R_{\lambda}$ the response function, and
$f_{\lambda}$ the received flux at earth (see for example Bessell et
al. 1998). We adopt the response curves of Bessell (1990) for the
standard Johnson-Cousins $UBVI$-system. The zero point calibration is
done on a model atmosphere of Vega (taken from the Kurucz homepage; a
file dated 17-april-1998, with parameters $T_{\rm eff}$ = 9550, $\log
g$ = 3.95, [m/H] = $-0.5$ and $v_{\rm turb}$ = 2 \ks), scaled to a
monochromatic flux at 5556 \AA\ of 3.54 $\times 10^{-9}$
erg/s/cm$^2$/\AA\ for a 0-magnitude star (Gray 1992).  On this scale
the $V$ magnitude of Vega is +0.03. The zero points in $V$, $(U-B)$,
$(B-V)$ and $(V-I)$ derived in this way are equal to, respectively,
$-21.082$, $-0.457$, 0.608 and 1.272 mag. In $V$ this is 0.01 brighter
than adopted by Fitzpatrick \& Massa (1999) and 0.018 fainter than
derived by Bessell et al. (1998), the difference with respect to the
former paper being probably due to the fact that we used a more recent
Kurucz model atmosphere for Vega. In $(U-B)$, $(B-V)$ and
$(V-I)$ the differences in the zero points in the sense ``this paper
minus Bessell et al. (1998)'' are $-0.003$, +0.002 and +0.004 mag,
respectively. To within the listed decimal figures we reproduce the
effective wavelengths of the $UBVI$ filters as given by Bessell et
al. (1998) for a A0{\sc v} star.

In case of the \M\ filter system, we obtained the response curves of
the total $JHK_{\rm s}$ system (that is, filter transmission, camera
response, dichroic reflectivity, typical atmospheric transmission)
from the \M\ Explanatory Supplement (Cutri et al. 2000 --
files labeled ``Total Response''). Using the same model for Vega as
before, effective wavelengths of, respectively, 1.228, 1.639, 2.152
$\mu$m, and $ZP$'s of, respectively, $-23.750$, $-24.857$ and
$-25.916$ mag in $JHK_{\rm s}$ are derived.

\subsection{The extinction curve}

The normalized UV extinction curve $k(\lambda-V)$ is described
by the following functional form (see for example, Fitzpatrick \&
Massa 1990):
\begin{equation}
k(\lambda-V) = c_1 + c_2 \; x + c_3/(\gamma^2+ (x -x_0^2/x)^2) + c_4 F(x)
\end{equation}
with $x = 1/\lambda$ in units of $\mu$m$^{-1}$, and with $F(x)$ = 0
for $x < 5.9 \, \mu$m$^{-1}$ and
\begin{equation}
F(x) = 0.5392 \; (x-5.9)^2 +0.05644  \; (x-5.9)^3
\end{equation}
for $x \ge 5.9 \mu$m$^{-1}$.
The choice of this particular form is based upon both physical
considerations (e.g. to fit the 2175 \AA\ feature with a Lorentz-like
profile) and empirical evidence (see Fitzpatrick \& Massa 1990, and
references therein for a discussion).

From Eq.~(1) and Eq.~(4) it is clear that possible degeneracies exist
in determining some of these parameters. First of all, $R$ and $c_1$
are dependent. In fact, G98a,b, fix the selective reddening at a value
of 3.1. Secondly, for a fixed $R$ value, there is a degeneracy between
$c_1$ and $\left(\frac{r_{\rm A}}{d}\right)^2$. With all other
parameters fixed, any change in $c_1$ can be countered by an
appropriate choice of $\left(\frac{r_{\rm A}}{d}\right)^2$ to give the
same model flux. It is therefore surprising to read in G98b that they
left free both $\left(\frac{r_{\rm A}}{d}\right)^2$ and $c_1, c_2,
c_3, c_4$. Although it is not explicitly stated in G98a which of the
parameters describing the extinction were left free, G98a refer to
Fitzpatrick \& Massa (1999), where it is noted that ``the linear terms
are combined'', probably referring to $c_1$ and $c_2$.

In fact, as was already noted by Fitzpatrick \& Massa (1990), there is
a strong correlation between $c_1$ and $c_2$. From Fitzpatrick 
(1999, his Eq.~(A2)):
\begin{equation}
c_1 = 2.030 -3.007 \times c_2
\end{equation}
with an estimated 1$\sigma$ dispersion of about 0.15. This was derived
for Galactic stars, but Misselt et al. (1999) show that stars in the
LMC fall on this relation as well.

There is another correlation that is used by us in order to further
constrain the UV extinction curve. From Fitzpatrick (1999, his Eq.~(A1)):
\begin{equation}
c_2 = -0.824 + 4.717/R
\end{equation}
with an estimated 1$\sigma$ dispersion of 0.12. 

G98a,b choose to fix the selective reddening at its mean Galactic
value of 3.1. However, even within our Galaxy there is a large spread
from 2.2 to 5.8 (quoted in Fitzpatrick 1999). This could also be true
for different lines-of-sight towards the LMC, and, in fact, Misselt et
al. (1999) derive values for $R$ towards LMC stars that range between
2.16 $\pm$ 0.30 and 3.31 $\pm$ 0.20.

In our fitting procedure we keep $R$ as a free parameter, fix $c_2$ at
its value determined by Eq.~(7) and fix $c_1$ at its value determined
by Eq.~(6). Other parameters left free are $x_0$, $\gamma$, $c_3$ and
$c_4$. The final error budget will take into account the scatter in
Eqs.~(6-7). We will also check the influence of fixing $R = 3.1$, and
leaving $c_2$ as a free parameter, as was probably done by G98a.

In the optical and near-infrared (NIR) wavelength range
($1.1 \le x \le 3.3 \mu$m$^{-1}$) the Galactic extinction curve by
O'Donnell (1994) is adopted, which is an improvement over the
Cardelli et al. (1989) one; it can be written as:
\begin{equation}
k(\lambda-V) = c_5 + \frac{ a(x) -1 + b(x)/R}{ 0.0014 + 1.0231/R}
\end{equation}
with $a(x)$ and $b(x)$ given in O'Donnell (1994).  The constant $c_5$
is introduced by us in order to join the UV and optical extinction
curve, but may also be thought of as to allow for a small difference
between the Galactic and LMC optical and NIR extinction curve. This
joining is done at 3.3 $\mu$m$^{-1}$, which is the blue edge of the
wavelength region where the extinction curve by O'Donnell (1994) is
valid. For given parameters $R, c_3, (c_1, c_2)$, the value of $c_5$
is determined (the value of $c_5$ does not depend on $c_4$). The
joining procedure allows only for continuity of the function
$k(\lambda-V)$ and not of its derivative. However, as we will see
below (Sect.~6.3) there is no change in slope at the joining point.
The value of $c_5$ is generally in the range $-0.1$ to $0.1$ (and in
particular $-0.006$ for the reference model 5) and this is a small
correction with respect to the second term in Eq.~(8) which typically
is 2.4 for $x = 3.3$.

\begin{table*}
\caption[]{Previous results}
\begin{tabular}{lccccl}  \hline
Author & $T_{\rm A}$ & [Fe/H] & $E(B-V)$ &  $\left( 
\frac{r_{\rm A}}{d}\right)^2$ & Extinction parameters \\ \hline

G98b & 22690 $\pm$ 300 & $-0.42 \pm 0.07$ & 0.086 $\pm$ 0.015 & $(1.99
\pm 0.13)\; 10^{-23}$$^{(a)}$  &  c$_1$, c$_2$, c$_3$, c$_4$ fitted;
$R$ fixed \\
G98a & 23000 $\pm$ 180 & $-0.45 \pm 0.06$ & 0.120 $\pm$ 0.009 & $(2.249
\pm 0.063)\; 10^{-23}$  &  unclear; $R$ fixed \\
\hline
\end{tabular}

Note: (a).  $\left(\frac{r_{\rm A}}{d}\right)^2$ derived from the
quoted distance of 49.8 $\pm$ 1.5 kpc. The error from the fitting
procedure alone is smaller than this.
\end{table*}

\begin{table*}
\caption[]{Fit results: main parameters}

\footnotesize

\begin{tabular}{crcccccl}  \hline
Model & \hspace{-6mm} $\chi^2_{\rm r}$ & $T_{\rm A}$ & [m/H] & $E(B-V)$ &  $\left( 
\frac{r_{\rm A}}{d}\right)^2$ & $R$ & Comments \\  
 & \hspace{-6mm}& (K) & & & (10$^{-23}$) & &  \\ \hline 
   & \hspace{-6mm}     &        &         &           &   & & \\
1  & \hspace{-6mm} 2.18 & 22982 $\pm$ 135 & $-0.36 \pm 0.04$ & 0.109 $\pm$ 0.006 & 2.12 $\pm$
0.06 & 3.17 $\pm$ 0.10 & Kurucz, standard errors (s.e.) \\
2  & \hspace{-6mm} 4.68 & 22865 $\pm$ 133 & $-0.11 \pm 0.04$ & 0.104 $\pm$ 0.006 & 2.16 $\pm$
0.07 & 3.16 $\pm$ 0.11 & Butler, s.e. \\
3  & \hspace{-6mm} 2.71 & 22486 $\pm$ 132 & $-0.12 \pm 0.04$ & 0.097 $\pm$ 0.006 & 2.23 $\pm$
0.07 & 3.39 $\pm$ 0.13 & Butler, s.e., without 3830 \AA\ point \\
4  & \hspace{-6mm} 2.16 & 22987 $\pm$ 135 & $-0.35 \pm 0.04$ & 0.110 $\pm$ 0.006 & 2.12 $\pm$
0.06 & 3.19 $\pm$ 0.12 & Kurucz, s.e., without 3830 \AA\ point \\
   & \hspace{-6mm}     &        &         &           &   & & \\
\hline
   & \hspace{-6mm}     &        &         &           &   & & \\
5  & \hspace{-6mm} 0.98 & 22932 $\pm$ 195 & $-0.38 \pm 0.07$ & 0.103 $\pm$ 0.007 & 2.05 $\pm$
0.07 & 3.06 $\pm$ 0.14 & Kurucz, Reference Model (R.M.)  \\

6  & \hspace{-6mm} 1.22 & 22467 $\pm$ 185 & $-0.13 \pm 0.06$ & 0.094 $\pm$ 0.007 & 2.18 $\pm$
0.09 & 3.31 $\pm$ 0.16 & Butler, final errors \\
7  & \hspace{-6mm} 0.98 & 22930 $\pm$ 195 & $-0.38 \pm 0.07$ & 0.103 $\pm$ 0.007 & 2.08 $\pm$
0.07 & 3.06 $\pm$ 0.14 & R.M. with $c_1$ larger by 0.15 \\
8  & \hspace{-6mm} 0.96 & 22962 $\pm$ 197 & $-0.44 \pm 0.07$ & 0.106 $\pm$ 0.007 & 2.13 $\pm$
0.08 & 3.24 $\pm$ 0.14 & R.M. with $c_2$ larger by 0.12 \\
9 & \hspace{-6mm} 0.98 & 22877 $\pm$ 194 & $-0.38 \pm 0.08$ & 0.102 $\pm$ 0.007 & 2.05 $\pm$
0.07 & 3.06 $\pm$ 0.14 & R.M. with $T_{\rm B}/T_{\rm A}$ = 1.010 \\
10 & \hspace{-6mm} 0.98 & 22978 $\pm$ 196 & $-0.38 \pm 0.07$ & 0.103 $\pm$ 0.007 & 2.04 $\pm$
0.07 & 3.03 $\pm$ 0.14 & R.M. with $\log g_{\rm A}$ = 3.563 \\
11 & \hspace{-6mm} 0.98 & 22949 $\pm$ 195 & $-0.38 \pm 0.07$ & 0.103 $\pm$ 0.007 & 2.04 $\pm$
0.07 & 3.05 $\pm$ 0.14 & R.M. with $\log g_{\rm B}$ = 3.605 \\
12 & \hspace{-6mm} 0.98 & 22931 $\pm$ 195 & $-0.38 \pm 0.07$ & 0.103 $\pm$ 0.007 & 2.03 $\pm$
0.07 & 3.06 $\pm$ 0.14 & R.M. with $\left(r_{\rm B}/r_{\rm A}\right)^2$ = 0.861 \\
13 & \hspace{-6mm} 0.93 & 22926 $\pm$ 200 & $-0.38 \pm 0.07$ & 0.102 $\pm$ 0.007 & 1.98 $\pm$
0.07 & 3.05 $\pm$ 0.15 & R.M., flux level FOS spectrum 3\% lower \\
14 & \hspace{-6mm} 1.04 & 22775 $\pm$ 192 & $-0.29 \pm 0.06$ & 0.100 $\pm$ 0.007 & 2.06 $\pm$
0.08 & 3.26 $\pm$ 0.15 & R.M., flux level FOS spectrum 1\% lower \\
   & \hspace{-6mm}     &        &         &           &   & &  beyond 3240 \AA\  \\
15 & \hspace{-6mm} 0.98 & 22954 $\pm$ 195 & $-0.38 \pm 0.07$ & 0.105 $\pm$ 0.007 & 2.07 $\pm$
0.07 & 3.09 $\pm$ 0.14 & R.M. with $(B-V) = -0.119$ \\
16 & \hspace{-6mm} 0.98 & 22958 $\pm$ 194 & $-0.37 \pm 0.06$ & 0.106 $\pm$ 0.007 & 2.09 $\pm$
0.07 & 3.12 $\pm$ 0.14 & R.M. with $(V-I) = -0.115$ \\
   & \hspace{-6mm}     &        &         &           &   & & \\
\hline
   & \hspace{-6mm}     &        &         &           &   & & \\
17 & \hspace{-6mm} 0.98 & 23052 $\pm$ 202 & $-0.33 \pm 0.06$ & 0.119 $\pm$ 0.010 & 2.24 $\pm$
0.11 & 3.34 $\pm$ 0.16  & no $(V-I)$ fitted, R.M. \\
18 & \hspace{-6mm} 1.31 & 22972 $\pm$ 188 & $-0.46 \pm 0.09$ & 0.076 $\pm$ 0.008 & 1.82 $\pm$
0.04 & 2.06 $\pm$ 0.13  & As 17, with {\sc tlusty} \\
19 & \hspace{-6mm} 1.21 & 22622 $\pm$ 202 & $-0.09 \pm 0.06$ & 0.113 $\pm$ 0.011 & 2.47 $\pm$
0.15 & 3.70 $\pm$ 0.17  & As 17, with Butler \\
20 & \hspace{-6mm} 1.02 & 22818 $\pm$ 201 & $-0.41 \pm 0.07$ & 0.093 $\pm$ 0.009 & 1.96 $\pm$
0.08 & 2.90 $\pm$ 0.20  & As 17, $(B-V) = -0.172 \pm 0.013$ \\
   & \hspace{-6mm}     &        &         &           &   & &  as constraint\\
21 & \hspace{-6mm} 0.95 & 23107 $\pm$ 214 & $-0.55 \pm 0.09$ & 0.102 $\pm$ 0.007 & 2.06 $\pm$
0.03 & 3.1  & As 17, $R$ fixed at 3.1\\
22 & \hspace{-6mm} 0.96 & 23178 $\pm$ 209 & $-0.55 \pm 0.09$ & 0.106 $\pm$ 0.007 & 2.07 $\pm$
0.03 & 3.1  & As 21, $V = 14.16 \pm 0.02$ \\
   & \hspace{-6mm}     &        &         &           &   & &  as additional constraint\\
   & \hspace{-6mm}     &        &         &           &   & & \\
\hline
\end{tabular}
\end{table*}

\section{Results}

In the {\it reference case} the 254 points of the FOS spectrum plus
$(B-V)$ and $(V-I)$ from UPW98 are used as the observational data-set.
Free parameters are the effective temperature of the primary, the
metallicity, $E(B-V)$, the scaling factor $\left(\frac{r_{\rm
A}}{d}\right)^2$, $R$, $x_0$, $\gamma$, $c_3$ and $c_4$. $c_1$ and
$c_2$ are derived from Eqs~(6-7); $c_5$ from the continuity of the
extinction and follows from the other parameters.  From the model
spectrum the values of $V$, $(U-B)$, $(B-V)$, $(V-I)$ and $JHK_{\rm
s}$ for the system are predicted.

Table~2 summarizes the results from G98a,b regarding the most
important fit parameters. The main difference is in the derived
distance, which resulted from using different (and more accurate)
$B,V$ photometry. Our results are listed in Tables.~3, 4 and 5 where,
respectively, the main fit parameters, the parameters describing the
UV extinction curve, and predicted broad-band colours are listed.

\begin{table*}
\caption[]{Fit results: parameters of the UV extinction law}


\begin{tabular}{ccccccccc}  \hline
Model & $x_0$ & $\gamma$ & c$_2$ & c$_3$ & c$_4$ \\
\hline 
1 & 4.68 $\pm$ 0.03 & 1.32 $\pm$ 0.11 & $-$ & 2.09 $\pm$ 0.40 & 0.64 $\pm$ 0.07 \\ 
2 & 4.69 $\pm$ 0.03 & 1.85 $\pm$ 0.15 & $-$ & 5.14 $\pm$ 1.13 & 0.75 $\pm$ 0.08 \\ 
3 & 4.71 $\pm$ 0.04 & 2.08 $\pm$ 0.18 & $-$ & 7.63 $\pm$ 1.75 & 0.97 $\pm$ 0.10 \\ 
4 & 4.68 $\pm$ 0.02 & 1.32 $\pm$ 0.11 & $-$ & 2.10 $\pm$ 0.40 & 0.65 $\pm$ 0.07 \\ 
\hline
5 & 4.67 $\pm$ 0.04 & 1.27 $\pm$ 0.16 & $-$ & 2.03 $\pm$ 0.58 & 0.66 $\pm$ 0.10 \\ 
6 & 4.70 $\pm$ 0.06 & 2.01 $\pm$ 0.26 & $-$ & 7.12 $\pm$ 2.42 & 0.97 $\pm$ 0.14 \\ 
7 & 4.67 $\pm$ 0.04 & 1.27 $\pm$ 0.16 & $-$ & 2.03 $\pm$ 0.59 & 0.66 $\pm$ 0.10 \\ 
8 & 4.67 $\pm$ 0.04 & 1.28 $\pm$ 0.17 & $-$ & 1.93 $\pm$ 0.57 & 0.58 $\pm$ 0.10 \\ 
9 & 4.67 $\pm$ 0.04 & 1.28 $\pm$ 0.16 & $-$ & 2.05 $\pm$ 0.59 & 0.66 $\pm$ 0.10 \\ 
10 & 4.67 $\pm$ 0.04 & 1.27 $\pm$ 0.15 & $-$ & 2.00 $\pm$ 0.58 & 0.66 $\pm$ 0.10 \\ 
11 & 4.67 $\pm$ 0.04 & 1.27 $\pm$ 0.15 & $-$ & 2.03 $\pm$ 0.58 & 0.66 $\pm$ 0.10 \\ 
12 & 4.67 $\pm$ 0.04 & 1.27 $\pm$ 0.15 & $-$ & 2.02 $\pm$ 0.58 & 0.66 $\pm$ 0.10 \\ 
13 & 4.67 $\pm$ 0.04 & 1.27 $\pm$ 0.16 & $-$ & 2.03 $\pm$ 0.60 & 0.66 $\pm$ 0.11 \\ 
14 & 4.69 $\pm$ 0.03 & 1.28 $\pm$ 0.15 & $-$ & 2.18 $\pm$ 0.61 & 0.79 $\pm$ 0.11 \\ 
15 & 4.67 $\pm$ 0.04 & 1.29 $\pm$ 0.15 & $-$ & 2.04 $\pm$ 0.59 & 0.65 $\pm$ 0.10 \\ 
16 & 4.67 $\pm$ 0.04 & 1.30 $\pm$ 0.14 & $-$ & 2.06 $\pm$ 0.59 & 0.65 $\pm$ 0.10 \\ 
\hline
17 & 4.68 $\pm$ 0.04 & 1.39 $\pm$ 0.17 & $-$ & 2.21 $\pm$ 0.62 & 0.63 $\pm$ 0.09 \\ 
18 & 4.57 $\pm$ 0.05 & 1.02 $\pm$ 0.18 & $-$ & 1.30 $\pm$ 0.46 & 0.35 $\pm$ 0.12 \\ 
19 & 4.74 $\pm$ 0.07 & 2.39 $\pm$ 0.32 & $-$ & 9.70 $\pm$ 3.48 & 0.92 $\pm$ 0.13 \\ 
20 & 4.67 $\pm$ 0.04 & 1.24 $\pm$ 0.16 & $-$ & 2.07 $\pm$ 0.61 & 0.70 $\pm$ 0.13 \\ 
21 & 4.66 $\pm$ 0.04 & 1.19 $\pm$ 0.15 & 0.88 $\pm$ 0.08 & 1.62 $\pm$ 0.50 & 0.47 $\pm$ 0.11 \\ 
22 & 4.66 $\pm$ 0.04 & 1.19 $\pm$ 0.15 & 0.86 $\pm$ 0.08 & 1.55 $\pm$ 0.47 & 0.46 $\pm$ 0.11 \\ 

\hline
\end{tabular}
\end{table*}

\begin{table*}
\caption[]{Fit results: colours }


\begin{tabular}{cccccccc}  \hline
Model & $V$ & $(U-B)$ & $(B-V)$  & $(V-I)$  & $J$    & $H$    & $K_{\rm s}$ \\
\hline
1  & 14.175 & $-0.869$ & $-0.124$ & $-0.095$ & 14.476 & 14.528 & 14.585 \\ 
2  & 14.187 & $-0.868$ & $-0.137$ & $-0.100$ & 14.509 & 14.542 & 14.612 \\ 
3  & 14.194 & $-0.866$ & $-0.142$ & $-0.102$ & 14.516 & 14.547 & 14.617 \\ 
4  & 14.174 & $-0.869$ & $-0.123$ & $-0.094$ & 14.475 & 14.526 & 14.583 \\ 
\hline
5  & 14.181 & $-0.870$ & $-0.130$ & $-0.105$ & 14.503 & 14.559 & 14.618 \\ 
6  & 14.198 & $-0.867$ & $-0.145$ & $-0.108$ & 14.532 & 14.565 & 14.636 \\ 
7  & 14.181 & $-0.870$ & $-0.130$ & $-0.104$ & 14.503 & 14.558 & 14.617 \\ 
8  & 14.178 & $-0.868$ & $-0.128$ & $-0.097$ & 14.482 & 14.534 & 14.590 \\ 
9  & 14.181 & $-0.870$ & $-0.130$ & $-0.105$ & 14.503 & 14.559 & 14.618 \\ 
10 & 14.181 & $-0.869$ & $-0.131$ & $-0.106$ & 14.506 & 14.562 & 14.622 \\ 
11 & 14.181 & $-0.870$ & $-0.130$ & $-0.105$ & 14.504 & 14.560 & 14.619 \\ 
12 & 14.181 & $-0.870$ & $-0.130$ & $-0.105$ & 14.504 & 14.559 & 14.618 \\ 
13 & 14.215 & $-0.870$ & $-0.131$ & $-0.105$ & 14.538 & 14.593 & 14.653 \\ 
14 & 14.194 & $-0.872$ & $-0.132$ & $-0.103$ & 14.510 & 14.563 & 14.621 \\ 
15 & 14.179 & $-0.869$ & $-0.128$ & $-0.101$ & 14.495 & 14.549 & 14.608 \\ 
16 & 14.178 & $-0.869$ & $-0.127$ & $-0.099$ & 14.489 & 14.542 & 14.600 \\ 
\hline
17 & 14.164 & $-0.866$ & $-0.115$ & $-0.078$ & 14.434 & 14.479 & 14.533 \\ 
18 & 14.221 & $-0.878$ & $-0.161$ & $-$ & $-$ & $-$ & $-$ \\ 
19 & 14.177 & $-0.862$ & $-0.128$ & $-0.075$ & 14.445 & 14.465 & 14.528 \\ 
20 & 14.191 & $-0.871$ & $-0.139$ & $-0.119$ & 14.540 & 14.600 & 14.662 \\ 
21 & 14.183 & $-0.870$ & $-0.132$ & $-0.104$ & 14.503 & 14.558 & 14.616 \\ 
22 & 14.179 & $-0.869$ & $-0.129$ & $-0.100$ & 14.492 & 14.546 & 14.604 \\ 

\hline
\end{tabular}
\end{table*}

\subsection{Models 1-4}

The first 4 models provide preliminary results. We have employed the
observational data of the reference case with the formal errors
derived from the FOS spectrum; both Kurucz and Butler model
atmospheres have been used.  It turns out that there is one wavelength
point (3830 \AA\ in the Kurucz wavelength grid) in the Balmer region
where the Butler model is severely off; as a test, fits with both the
Kurucz and Butler spectra have been performed excluding this observed
wavelength point (models 3 and 4), in order to better compare the 2
sets of theoretical spectra. Also in this case the Kurucz models
provide the lower (reduced) $\chi^2$, and therefore they will be
adopted from now on as the standard theoretical spectra. Morever, the
results for the fitted parameters obtained with the Kurucz spectra are
completely insensitive to the inclusion or not of this wavelength
point.

It is interesting to note that the [m/H] value found with Butler data
is higher than with the Kurucz ones. This is related to the fact that
the adopted solar Fe abundance adopted by Butler is lower (more in
agreement with the currently accepted value, see previous discussion)
than in the case of Kurucz ATLAS 9 models. The difference of 0.21 dex
in the [m/H] value derived from the fitting is exactly the same as the
difference in the adopted solar iron content between the two sets of
model atmospheres.  This is consistent with the claim by Fitzpatrick
\& Massa (1999) that the Fe abundance is the main parameter
responsible for the UV spectral signature of the total metal content.
In this respect it is also worth pointing out that in Table 2 we
summarize the G98a and G98b results showing their derived [Fe/H]
abundances.  We believe that these [Fe/H] abundances are derived
applying (as suggested by Fitzpatrick \& Massa 1999) a correction to
the derived [m/H] values taking into account the difference between
the ATLAS 9 solar iron abundance and the Grevesse \& Noels (1993) one;
this should be taken into account when comparing results in Table 2
with the results listed in Table 3.

A remark must be also made regarding the high values of $\chi^2$
obtained for models 1-4 
which would indicate that they are not a good representation of the
observations. G98a find a reduced $\chi^2$ ``close to one'' also using
Kurucz spectra. It seems likely that the difference is not due to the
theoretical spectra, but rather to the assigned observational
errors. In fact, the errors associated with the FOS spectrum are
formal errors only, and do not include the error due to sky and
background subtraction and flat field errors (see the {\sc fos}
handbook). Furthermore, the FOS spectrum was binned. However, the
wavelength points are not independent as each point was observed by
several diodes in the FOS spectrograph. The errors in the FOS spectrum
are therefore underestimates, but it is not simple to calculate the
true error for each wavelength point. We therefore chose to scale all
errors by a factor 1.5, to give a reduced $\chi^2$ of approximately
1.0 for the standard model. This also allows a direct comparison of
the derived errors with G98a.

\subsection{Models 5-16}

Model 5 represents the {\it reference model}, that is, the reference
case with Kurucz spectra plus the scaling of the errors described
above.  The validity of scaling the errors is demonstrated by the fact
that the values for the parameter are essentially unchanged (within
1$\sigma$) with respect to model 1. Only the errors on the derived
quantities are increased; this probably makes these errors a
conservative estimate of the real errors.  Comparison of model 5 with
the previous results in Table~2 shows that the results are in better
agreement with G98b than with G98a. However, since in our case
different constraints on the reddening law are used, and also the
$(V-I)$ is added to the wavelength range to be fitted with theoretical
spectra, it is difficult to compare the results directly. Such an
attempt will be made with model 21.

In model 6 a fit is performed as for model 5 (but without the FOS
spectrum point at 3830 \AA ) using the Butler model
atmospheres. Significant differences are derived for some of the
extinction parameters as well as for the distance (the scaling factor
$\left(\frac{r_{\rm A}}{d}\right)^2$) and the effective temperature of
the primary component.  The resulting $\chi^2$ is however larger
than when using ATLAS 9 spectra.

Models 7-16 show the influence of 9 sources of errors on the results
of model 5. In models 7 and 8, $c_1$ and $c_2$ are increased by their
respective 1$\sigma$ uncertainty; models 9-12 investigate the
uncertainty in the binary parameters, as derived from the radial
velocity and lightcurve solution, while model 13 investigates the
influence of the overall absolute photometric calibration on the
distance. This calibration error nominally is 3\% according to the
{\sc fos} handbook, due to the uncertainties on the White Dwarf
model--based reference flux system and several other instrumental and
observational issues. For individual observations this error may be
larger.  Remarkably enough, the error on the derived parameters
appears dominated by the fitting procedure and not by errors on the
constraints.  In model 14, the spectrum beyond 3240 \AA\ is decreased
by 1\% to simulate possible offsets between the different parts of the
{\sc fos} spectrum, while in models 15 and 16 the constraints on
$(B-V)$ and $(V-I)$ are changed by 0.01 mag to simulate the absolute
photometric error.  The derived parameters are in agreement, within
the respective error bars, with the results of model 5.

\subsection{Models 17-22}

In models 17-19, as a test, we consider the reference case plus
rescaled errors, but excluding from the fit the observed $(V-I)$
value.  We employed the three different sets of theoretical spectra
described in Section 4.1 because in this case the Hubeny {\sc tlusty}
spectra cover all the relevant wavelength range.  It is interesting to
notice for model 17 that the exclusion of $(V-I)$ from the fit
produces a difference at the level of 1.2$\sigma$ for the derived
values of $E(B-V)$, $R$ and $\left(\frac{r_{\rm A}}{d}\right)^2$.  In
the hypothesis that the ATLAS 9 models are a correct representation of
the actual stellar spectra, at least for this kind of stars, this
occurrence implies that there is information in the $(V-I)$ colour
which must be taken into account when deriving the stellar properties.
In this respect it is unfortunate that there are no data at longer
wavelengths suited to be used in the fitting procedure.

The same fit with Butler spectra (model 19) provides a 
higher $\chi^2$ with respect to Kurucz data. The derived distance is
extremely short, significantly different from the reference model.
The fit using {\sc tlusty} spectra (model 18) provides an even higher
$\chi^2$.  In particular, the fit in the 1450-1650 \AA \ region is
unsatisfactory. {\sc tlusty} spectra provide by far the longest
distance and lowest reddening, metallicity (especially considering the
fact that the solar Fe content in {\sc tlusty} spectra is 0.17 dex
lower than in the ATLAS 9 ones and 0.04 dex larger than for Butler
models), effective temperature and $R$.

So far, the broad-band photometry of UPW98 was used because it
included $(V-I)$. Model 5 fits well these constraints: $(B-V)$ =
$-0.135$ versus the observed value of $-0.129 \pm 0.015$, and $(V-I)$
= $-0.110$ versus the observed value of $-0.125 \pm 0.015$, while $V$
= 14.187 is predicted within $1.4\sigma$ of the observed value 14.16
$\pm$ 0.02.  Model 20 is similar to model 17, with the only exception
than in this case we used the observed $(B-V)$ = $-0.172 \pm 0.013$
from Nelson et al. (2000) as constraint. This leads to a longer
distance to the binary, the parameter $\left(\frac{r_{\rm
A}}{d}\right)^2$ being 1.4$\sigma$ lower than in the case of model 5.
The $\chi^2$ is only slightly higher than in the case of model 17, but the
observed $V$ and $(B-V)$ values considered in this case are more
poorly reproduced with respect to model 5 or model 17.  In particular,
the predicted $(B-V)$ value is about 2$\sigma$ different from the
observed one.

Model 21 is probably the closest to what is our understanding of what
G98a did, that is, model 6 but fixing $R=3.1$ and not using $(V-I)$ as
constraint. However, reddening and distance differ significantly from
those quoted in G98a. In particular, we get a longer distance and
lower reddening; we were unable to trace the reasons for this
difference. While the values of the derived parameters, with the
exception of the metallicity, are consistent within 1$\sigma$ with the
results from the reference model, it is remarkable the difference with
respect to the distance derived from model 17.  The only constraint
changed between these two cases is that in model 21 we kept $R$ fixed
at 3.1, while in model 17 $R$ was a free parameter.

Model 22 is as model 21, with the inclusion of the observed 
$V$ = 14.16 $\pm 0.02$ as additional constraint to the fit. The derived
parameters are almost exactly the same in both models.



\begin{figure*}
\centerline{\psfig{figure=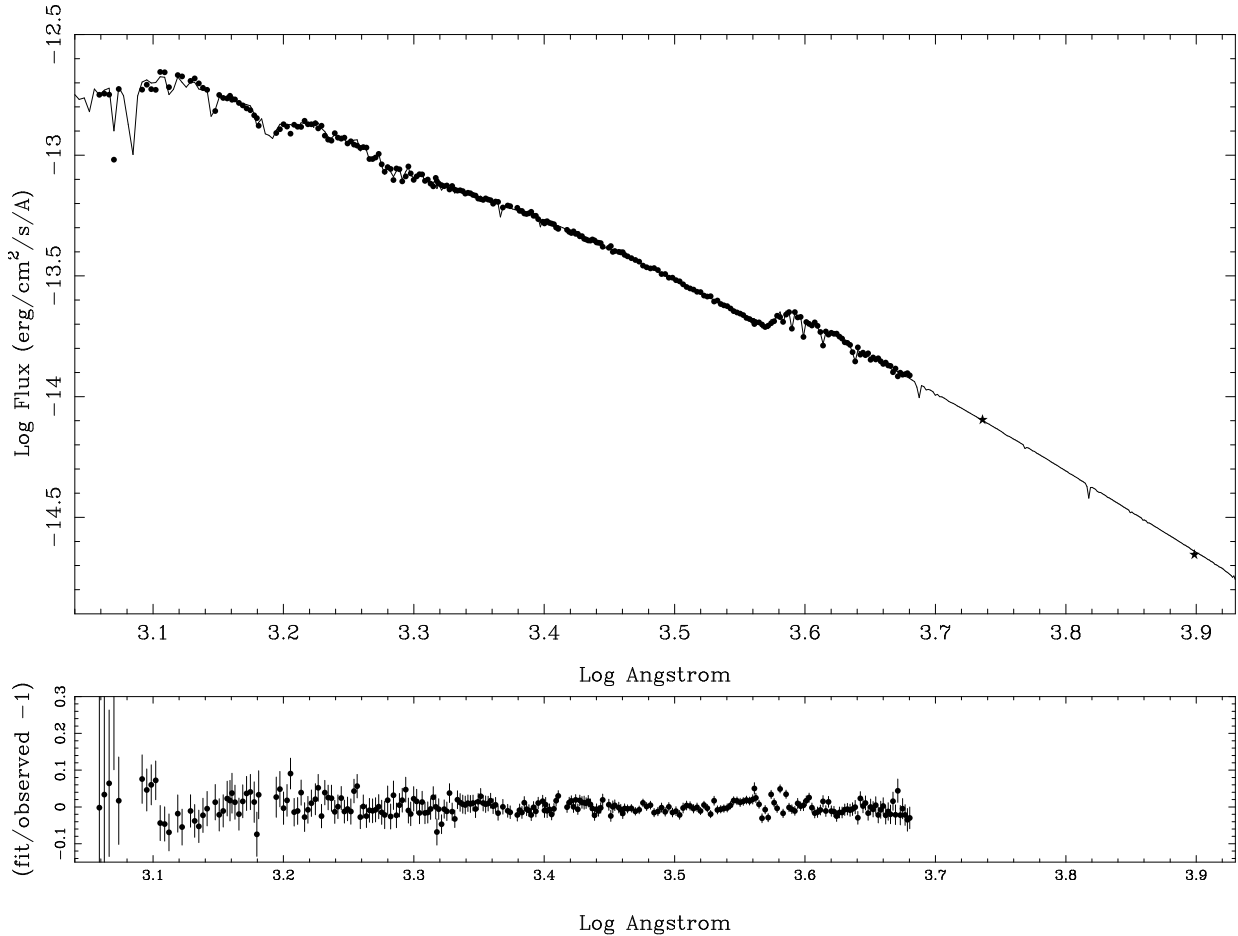,width=16.5cm}}
\caption[]{Best fit to the observed data (model 5). In the lower panel
the residuals are plotted, with the error in the individual
observation. $V$ and $I$ are plotted at the effective wavelenghts
calculated from folding the energy distribution with the filter
response curves.}
\end{figure*}

\section{Discussion}

We will now discuss the results obtained in the previous section, with
particular emphasis on the consistency of the derived parameters for
the reference model.

\subsection{The distance to HV 2274}

As previously explained, our reference model is model 5, which has
the maximum wavelength coverage -- from the UV to the near infrared --
of the HV2274 spectrum, and makes use of the set of theoretical
spectra which provides the lower $\chi^2$ value.  In Fig.~1 and 2 we
show, for model 5, the fit of the theoretical spectra to the FOS one,
and the resulting UV and optical normalized extinction curve. 

The final value for $\left( \frac{r_{\rm A}}{d}\right)^2$ is that of
model 5. The final error estimate comes from the formal error in the
model 5 result, added in quadrature to the squared differences of the
parameter values of model 5 with those of models 7-16, and the
additional error due to the error in  $r_{\rm A}$. The result is
$\left( \frac{r_{\rm A}}{d}\right)^2$ = (2.05 $\pm$ 0.12)
10$^{-23}$. This corresponds to a linear distance of 49.10 $\pm$ 1.44
kpc, or a true distance modulus of 18.46 $\pm$ 0.06.

\subsection{The distance to the center of the LMC}

To obtain the distance to the center of the LMC, the location of the
binary with respect to the LMC center has to be taken into account.
The geometry of the LMC can be described by an inclined disk, and one
therefore has to consider the distance from the plane of that disk to
the plane of the sky through the center of the LMC at the position of
the binary, and the fact that the binary may be behind or in front of the
plane of the disk.  G98a assume that HV 2274 is 1.1 kpc behind the
center of the LMC, based on the parameters of Schmidt-Kaler \&
Gochermann (1992).

Various authors have described the geometry of the LMC with a thin
disk and derived the position angle ($\theta$) of the line-of-nodes
and inclination angle ($i$). The relevant coordinate transformation to
go from observed right ascension and declination to a rectangular
coordinate system in the plane of the sky, and to a similar one
rotated by $\theta$ and $i$ are given in Weinberg \& Nikolaev (2000).
Table~6 gives estimates of the distance of the LMC plane to the center
of the LMC for various accurate estimates of $\theta$ and $i$, using
the coordinate system by Weinberg \& Nikolaev (2000). Because of the
different orientations and definition of positive inclination the
values in the table may differ by 90\degr\ with respect to the values
quoted in the original reference.

The mean of these 5 determinations is 0.88 kpc; the average of the
highest and lowest value is 0.89 kpc and the median is 0.82 kpc. The
adopted difference in distance is 0.9 $\pm$ 0.3 kpc\footnote{As an
aside we did the same for SN 1987A. The shortest distance (SN 1987A -
LMC Center) is $-0.64$ kpc [for the parameters of Schmidt-Kaler \&
Gochermann (1992)], and the largest is $-0.15$ kpc [for the parameters
of Groenewegen (2000)]. Based on all 5 determinations, the LMC plane
at the location of SN1987A is 0.4 $\pm$ 0.2 kpc in front of the LMC
center.}.  The error does not come from the internal errors of each of
the determinations, but from the spread among the values itself, and
the difference between the lowest and highest value has been assumed
to correspond to 3$\sigma$.

It is unknown if HV 2274 is in front or behind the LMC plane. The
vertical scale height of the LMC disk is small however, between 100
and 300 pc (see the discussion in Groenewegen, 2000). The largest
value is taken here, and added to the error mentioned above, to give
the final result that HV 2274 is located 0.9 $\pm$ 0.5 kpc behind the
LMC center.  Considering a DM to HV 2274 of 18.46 $\pm$ 0.06, the DM
to the LMC center 18.42 $\pm$ 0.07.

\begin{table}
\caption[]{The distance (HV 2274 $-$ LMC center) }

\begin{tabular}{cccccccc}  \hline
$\theta$ & $i$     & Reference & $\Delta$ \\
 (\degr) & (\degr) &           & (kpc)    \\
\hline
 258     & 38  & Schmidt-Kaler \& Gochermann (1992) & 1.30 \\ 
 258     & 33  & Feitzinger et al. (1977)           & 1.08 \\
 261     & 25  & Weinberg \& Nikolaev (2000)        & 0.82 \\
 296     & 18  & Groenewegen (2000)                 & 0.75 \\
 232     & 29  & Martin et al. (1979)               & 0.47 \\
\hline
\end{tabular}
\end{table}

\subsection{The extinction curve towards HV 2274}

From model 5 we get $x_0 = 4.67 \pm 0.04$, $\gamma = 1.27 \pm 016$,
$c_3 = 2.03 \pm 0.58$, $c_4 = 0.66 \pm 0.10$ and $R = 3.06 \pm 0.14$
(where the errors are the internal errors due to the fitting procedure
only); from Eqs.~(6) and (7) one obtains $c_1 = -0.13 \pm 0.21$ and
$c_2 = 0.72 \pm 0.07$.  These values are consistent with the
corresponding quantitites recently determined by Misselt et al.~(1999)
for their 'LMC-Average Sample' of stars.  Figure 2 shows the resulting
normalised UV and optical extinction curve.

\subsection{The reddening towards HV 2274}

The value of $E(B-V)$ obtained from model 5 is $E(B-V)$ = 0.103 $\pm$
0.007.  It is important to notice that this value is different from
the one derived by G98b ($E(B-V)$ = 0.120 $\pm$ 0.009) and from the
determination by UPW98 ($E(B-V)$ = 0.149 $\pm$ 0.015) who used
colour-colour $(U-B)-(B-V)$ relationships.  Since we are using the
$(B-V)$ (and $(V-I)$) photometric data by UPW98 as constraint for the
spectrum fit we investigated the possibility that the reddening
derived from the spectrum fit is inconsistent with HV2274 broad band
photometry.

We followed the same procedure by UPW98 that is, to consider a local
standard $(U-B)-(B-V)$ sequence of stars with the same spectral type
as HV 2274, and derive the reddening from the displacement of the
position of HV 2274 with respect to the standard sequence, for an
assumed $E(U-B)/E(B-V)$ ratio. We used the same standard colour-colour
sequence employed by UPW98 (for B-stars of luminosity class III), and
a ratio $E(U-B)/E(B-V)$ derived from our adopted extinction law using
the value of $R$ derived from the fit (which is by the way very close
to the standard value of 3.1).  By computing appropriate stellar
models (using the same input physics and colour transformations as in
Salaris \& Weiss 1998 and a scaled-solar metal distribution) for solar
metallicity and the metallicity derived from model 5, with masses
around 12$M_{\odot}$, we verified that the location of B stars on the
$(U-B)-(B-V)$ plane does not vary in this metallicity range.

As for the $(B-V)$ colour of HV 2274 we used the value by UPW98, while
for $(U-B)$ we derived the value from the FOS spectrum (see Section
4.2), since it covers the wavelength region spanned by this colour
index.  The $(U-B)$ colour derived from the spectrum is different from
the value observed by UPW98.  The FOS spectrum provides $(U-B)$ =
$-$0.836 $\pm$ 0.006, while UPW98 measured $(U-B)$ = $-$0.905 $\pm$
0.04. Nelson et al.~(2000) measured $(U-B)$ = $-0.793$ $\pm$ 0.031
which is closer (but still inconsistent at the 1$\sigma$ level) to the
value derived from the FOS spectrum. From the colour-colour diagram we
get $E(B-V)$ = 0.115 $\pm$ 0.015 which is consistent, within 1$\sigma$, with the
value derived from model 5.

\subsection{The metallicity of HV 2274}

The value of [m/H] from the reference model, taking the formal error
from the fit plus the external errors from models 7-16 is $-0.38 \pm
0.12$. This is relative to the adopted iron abundance of 7.67 in the
Kurucz models. This implies that the metallicity relative to the
currently favoured solar iron abundance of 7.51 (Grevesse \& Noels
1993) is [Fe/H] = $-0.22 \pm 0.12$. Model 6 computed using Butler 
atmospheres has [m/H]
= $-0.13$ relative to a solar abundance of 7.46,
[Fe/H] = $-0.18$ relative to the solar abundance of Grevesse \& Noels.
These values are in good agreement with each other, but significantly
more metal rich than the value of $-0.42 \pm 0.07$ to $-0.45 \pm 0.06$
found by G98a,b.

Observationally there are few direct iron abundance determinations for
hot and young stars in the LMC. Haser et al. (1998) derived a
metallicity of $-0.3$ and $-0.1$ dex for an O3{\sc iii} and O4{\sc i}
star in the LMC. Korn et al. (2000) determined the iron abundance in 5
non-supergiant B-stars, the average value being $-0.42 \pm 0.15$. Both
the values found by G98a,b and ours are consistent with these
determinations.

\subsection{The \M\ data}

As mentioned in Sect.~3 HV 2274 has been detected in the \M\ $JHK_{\rm
s}$ infrared survey. The observation date is JD = 2451111.689. From
the ephemeris in Watson et al. (1992) a primary eclipse is predicted
at JD = 2451111.697 $\pm$ 0.006, taking into account the error in the
time determination of the reference primary eclipse and the period. The
shift of 0.008 $\pm$ 0.006 days, or 0.0014 in phase is negligible,
also in light of the fact that phase shifts of up to 0.05 occur due to
apsidal rotation (Watson et al. 1992).

In the optical the magnitude difference between out-of-eclipse and
primary eclipse are 0.72, 0.71 and 0.70 mag in $BVI$, respectively
(Watson et al. 1992). Naively, one might therefore expect a magnitude
difference of about 0.69 in $J$.  The difference between the observed
\M\ $J$ and the model predictions is (15.152 $\pm$ 0.057) $-$ (14.503
$\pm$ 0.049) = 0.65 $\pm$ 0.08, where the error in the model comes
from summing up all the error terms from models 6-17. This is in good
agreement with the value extrapolated from the magnitude difference
between out-of-eclipse and primary eclipse measured in the optical.
However, as emphasised later, out-of-eclipse NIR photometry would be
important in further constraining the parameters of the model.

\begin{figure}
\centerline{\psfig{figure=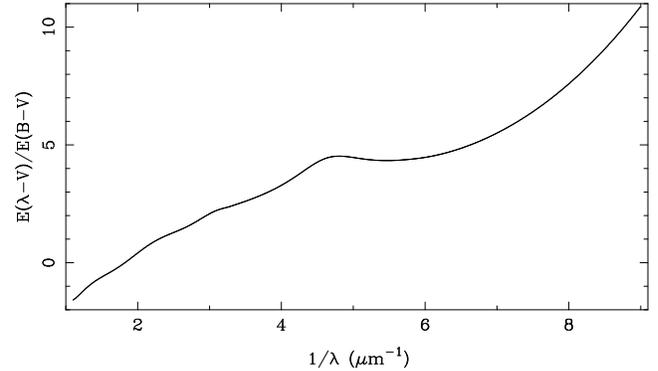,width=8.5cm}}
\caption[]{UV and optical normalised extinction curve for the 
parameters of model 5.}
\end{figure}

\subsection{Final remarks}

In this paper we have described and analyzed in detail the method used
by G98a,b to derive the distance to the eclipsing binary HV 2274 in
the LMC. We used various sets of theoretical spectra and broad band
photometric data in the fitting procedure outlined in Section 4, and
we found that Kurucz ATLAS 9 spectra best reproduce the observed
spectrum of HV 2274.  The selection of the wavelength range to be
covered by the spectrum fit and the constraint on $R$ play also a role
in determining the outcome of the fitting procedure. We are now going
to comment briefly about this point.

In the reference model we fitted HV 2274 FOS data plus $(B-V)$ and
$(V-I)$ colours (from UPW98) using ATLAS 9 spectra, constraining the
parameters $c_1$ and $c_2$ of the UV extinction law according to
Equations (6) and (7) and keeping $R$ as a free parameter.  The use of
$(V-I)$ is dictated by our desire to use all available information
about the spectral energy distribution of HV 2274.  The homogeneous
$UBVI$ photometry by UPW98 makes it possible to cover all the spectral
range from UV to near-infrared.

Neglecting $(V-I)$, that means, fitting a smaller wavelength range,
induces a decrease by 0.10 mag in the distance modulus derived from
the fitting procedure.  If we consider the ATLAS 9 models as an
accurate reproduction of the 'real' spectra of B stars, this
difference can be ascribed to the fact that there is information about
the distance contained in the $(V-I)$ colour, and therefore it must be
included in the fitting procedure.  Conversely, this result could also
imply that ATLAS 9 spectra are inconsistent with observations at the
longer wavelengths or that broad-band photometry and FOS spectrum are
not homogeneously calibrated. In this respect, one should keep in mind
that the $(U-B)$ value derived from the FOS spectrum is inconsistent
with the $(U-B)$ determined by UPW98 and Nelson et al.~(2000) and also
that the $(U-B)$ values derived from the best-fit ATLAS 9 spectrum are
systematically different by $\sim$0.03 mag with respect to the FOS
colour (see Table~5).

In case of not considering $(V-I)$ for the spectrum fitting, keeping
$R$ fixed increases sensibly (0.09 mag) the derived distance modulus
with respect to the case of having $R$ determined by the fitting
procedure, as shown by comparing the outcome of models 21 and 17.
However, it is not clear why one should fix the value of $R$ a priori,
since it is known that it is subjected to variations within the LMC
and within our galaxy.  Finally, the use of the $(B-V)$ by Nelson et
al.~(2000) together with the FOS spectrum produces a distance modulus
even larger (0.16 mag larger than the case of using UPW98 data).  The
quality of the fit is only marginally lower, and the reddening
compares well with the value ($\sim$ 0.08) one would derive from the
colour-colour diagram considering the Nelson et al.~(2000) $(B-V)$ and
the FOS $(U-B)$.

A last comment is that it is very unfortunate that the \M\ infrared
data happens to be taken during eclipse. The difference in predicted
$JHK$ between the different models is up to 0.15 mag (see
Table~5). Therefore, accurate NIR photometry or even NIR spectroscopy
at the 1\% level is expected to give valuable additional
constraints. \\

The conclusion is that the use of eclipsing binaries as distance
indicators is powerful, but that a photometrically well calibrated
data set covering a large wavelength region is essential. Furthermore,
the discrepancies among the different theoretical model atmospheres is
worrying and need further investigation.

%

\acknowledgements{

It is a pleasure to thank Keith Butler (Munich Observatory) and Ivan
Hubeny (Goddard Space Flight Center) for providing us with model
atmospheres tailored to our specific needs, Ignasi Ribas for
many helpful remarks, and Phil James for a preliminary reading 
of the manuscript. 

This work is based on observations made with the NASA/ESA Hubble Space
Telescope, obtained from the data archive at the Space Telescope
Science Institute. STScI is operated by the Association of
Universities for Research in Astronomy, Inc. under NASA contract NAS 5-26555.

This publication makes also use of data products from the Two Micron
All Sky Survey, which is a joint project of the University of
Massachusetts and the Infrared Processing and Analysis
Center/California Institute of Technology, funded by the National
Aeronautics and Space Administration and the National Science Foundation.

}

{}

\end{document}